\documentclass[prb,letterpaper,aps,10pt,superscriptaddress,twocolumn]{revtex4-1}
\usepackage{amsmath, upgreek, amssymb, graphicx, paralist, gensymb, subfigure}
\usepackage[colorlinks, linkcolor=blue, citecolor=blue, urlcolor=blue]{hyperref}

\newcommand{\eref}[1]{Eq.~(\ref{#1})}

\newcommand{\fref}[1]{Fig.~\ref{#1}}

\newcommand{\rmd}{\text{d}}
\newcommand{\ket}[1]{\lvert#1\rangle}
\newcommand{\bra}[1]{\langle#1\rvert}

\newcommand{\abs}[1]{\lvert#1\rvert}

\newcommand{\im}[1]{\text{Im}\!\left\{#1\right\}}
\newcommand{\re}[1]{\text{Re}\!\left\{#1\right\}}
\DeclareMathOperator{\tr}{Tr}
\DeclareMathOperator{\Tr}{Tr}

\makeatletter
\newlength \figurewidth
\renewcommand\@thesubfigure{}
\makeatother

\begin{document}

\title{Optomechanical interface for probing matter-wave coherence}

\date{\today}

\author{Andr\'e Xuereb}
\affiliation{Centre for Theoretical Atomic, Molecular and Optical Physics, School of Mathematics and Physics, Queen's University Belfast, Belfast BT7\,1NN, United Kingdom}
\affiliation{Department of Physics, University of Malta, Msida MSD\,2080, Malta}
\author{Hendrik Ulbricht}
\affiliation{School of Physics and Astronomy, University of Southampton, Southampton SO17\,1BJ, United Kingdom}
\author{Mauro Paternostro}
\affiliation{Centre for Theoretical Atomic, Molecular and Optical Physics, School of Mathematics and Physics, Queen's University Belfast, Belfast BT7\,1NN, United Kingdom}

\begin{abstract}
\begin{center}Please address all correspondence to:\ \href{mailto:andre.xuereb@gmail.com}{andre.xuereb@gmail.com}\end{center}
\par
We combine matter-wave interferometry and cavity optomechanics to propose a coherent matter--light interface based on mechanical motion at the quantum level. We demonstrate a mechanism that is able to transfer non-classical features imprinted on the state of a matter-wave system to an optomechanical device, transducing them into distinctive interference fringes. This provides a reliable tool for the inference of quantum coherence in the particle beam. Moreover, we discuss how our system allows for intriguing perspectives, paving the way to the construction of a device for the encoding of quantum information in matter-wave systems. Our proposal, which highlights previously unforeseen possibilities for the synergistic exploitation of these two experimental platforms, is explicitly based on existing technology, available and widely used in current cutting-edge experiments.
\end{abstract}

\maketitle

The behaviour of Nature at the atomic scale is, by all accounts, described exceedingly well by quantum mechanics. Quite differently, there is still an active debate surrounding the question of whether the same is true of the Nature at the ``meso-scale''~\cite{Adler2009,Pepper2012,Bassi2013}, i.e., the behaviour of objects in-between the micro-scale atomic world and the macro-scale `classical' world of everyday experience. Several ideas have been put forward in an attempt to explain the apparent non-existence of quantum superposition states of the mesoscopic systems. Indeed, one ubiquitous pitfall is the interaction of such systems with their environment, for it is far easier to isolate a single atom from its environment almost perfectly than it is a mesoscopic system. Current experimental evidence cannot rule out, however, less mundane explanations, e.g., continuous spontaneous localisation~\cite{Home1996,Nimmrichter2011}, where quantum dynamical equations are augmented by terms that cause meso-scale superposition states to `collapse' of their own accord, or gravitational collapse~\cite{Karolyhazy1966,Diosi1984,Penrose1996,Blencowe2013}, where it is the difference in the gravitational potential between the states forming a superposition that causes this collapse. In an attempt to answer this question, two disjoint experimental programmes are being followed. The `top-down' approach typical of optomechanics (OM)~\cite{Meystre2013,Aspelmeyer2013} investigates mechanical systems of decreasing effective size in order to investigate the quantum mechanical behaviour of their dynamics. The `bottom-up' approach, as pioneered in molecular matter-wave interferometry experiments~\cite{Hornberger2012}, exposes the quantum-mechanical behaviour of molecules of ever-increasing mass.\par
These two approaches, involving molecules with masses up to $\sim10^{-23}$\,kg~\cite{Gerlich2011} and mechanical oscillators whose effective masses can be as low as $\sim10^{-15}$\,kg~\cite{Eichenfield2009b} are separated by eight orders of magnitude. Our principal aim in this article is to provide a bridge between these two approaches. Extending on the role played by OM systems as effective interfaces between photons of vastly differing frequencies~\cite{Bagci2013}, we introduce the concept of OM as an interface between electromagnetic and matter-wave fields. By putting together state-of-the-art experimental technologies, we show that the system we propose is able to coherently interface these two vastly different physical systems. An in-depth study of particular realisations of our scheme lies outside the scope of this article, which aims to provide a proof of principle showing that signatures of the operation of this coherent interface will be experimentally observable.\\
Aside from providing an interface between electromagnetic and matter waves, the system we propose provides a basis for integrating matter waves into quantum networks~\cite{Kimble2008}. It therefore extends the reach of such networks in a completely new direction and introduces the tantalising possibility of indirect interactions between two matter waves separated in space and time.

\section*{Results}
\begin{figure}[th]%
\includegraphics[width=1.0\figurewidth]{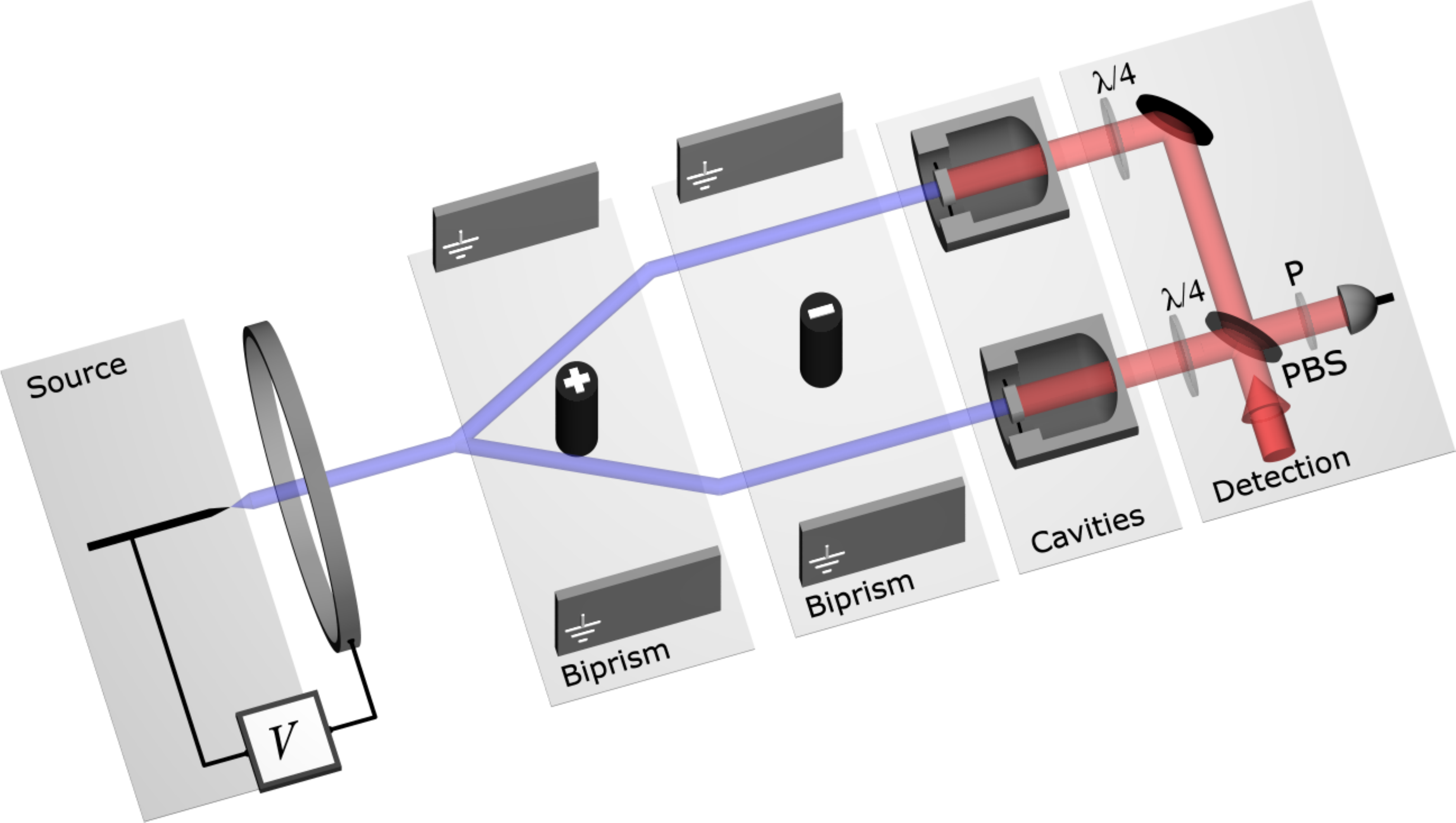}
 \caption{\textbf{Schematic diagram of the system discussed.} Ions from a point source (blue path) are accelerated through a potential $V$. The resulting wavefronts are split using two biprisms, and the two arms sent to two optomechanical cavities, shown in half-section. The signal is read out using standard optical techniques. Here we show a setup that erases `which-way' information. Light (red path) is split at a polarising beamsplitter (PBS) and passes through two $\lambda/4$ waveplates on its way to the two cavities. At the beamsplitter, the reflected light therefore takes the opposite route. A polariser `P' then selects a polarisation at $45$\degree, mixing the two signals.}%
 \label{fig:System}%
\end{figure}%
\noindent\textbf{Description of the system.} We begin by describing the physical system that forms the basis of our discussion, illustrated in \fref{fig:System}. Let us emphasise at this stage that this system serves as a convenient prototype for the paradigm we introduce, and we do not exclude the existence of other, possibly more effective, interfaces.\\
A beam of ions (`particles') is produced by a point-like field-emission source providing the sufficient transverse coherence of the particle beam to be diffracted. The particles are then sent through an acceleration stage to control their momentum and to narrow their velocity distribution. The matter-wavefront is split into distinct beams by a biprism to form the two arms of an interferometer. Such ion beam generation and manipulation is based on existing technology~\cite{Hasselbach2010}. Transverse coherence is maintained between the two separated wavefronts, which are subsequently incident onto two identical vibrating micro-mechanical mirrors. The latter act as the end-mirrors of two optical cavities, in turn driven by light, which are used for the read-out of the mechanical states. Optical tomography of the mechanical state of the mirrors is then performed, e.g., using the pulsed optomechanics scheme proposed in Ref.~\onlinecite{Vanner2011b} and demonstrated recently~\cite{Vanner2012b}. Recent advances~\cite{Narevicius2007,Narevicius2008,Raizen2009} in the control of atomic beams raise the question of whether it is possible to reproduce our scheme using neutral atoms or molecules rather than ions, since potentially problematic interparticle interactions are much weaker for the former. Whereas in principle this is possible, the high momentum required of the individual particles is significantly easier to obtain with charged particles, as is the wavefront-splitting procedure necessary in our scheme.\par
Before we move on to the mathematical description of our system, let us first provide some physical motivation for our model. The state of $N$ particles before collision with the mirrors is a superposition of states with $r$ particles in one arm of the interferometer and $N-r$ particles in the other arm. The use of this particular state is a first step, and we do not claim overall optimality for the scheme put forward herein. Each pure state making up the superposition contains no information other than the number of particles in each arm; the internal state of each particle is not relevant to our study, since it is destroyed upon collision and the interaction model that we seek and address is insensitive to such degrees of freedom. Any interference effects therefore arise at the single-particle level, as is the case in all molecular matter-wave interferometry experiments performed so far~\cite{Hornberger2012}. This information is transferred to the mirrors upon collision.\\
\begin{figure}[th]%
\includegraphics[scale=1.25]{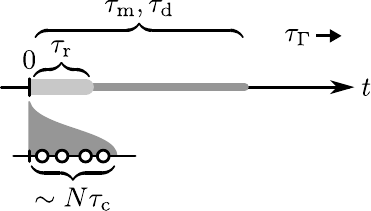}
 \caption{\textbf{Timescales of the problem.} The collisions happen over a time $N\tau_\mathrm{c}$, which is much smaller than the readout time $\tau_\mathrm{r}$. In turn, $\tau_\mathrm{r}$ must be much smaller than the decoherence timescale ($\tau_\mathrm{d}$) and the timescale for the decay of mechanical excitations of the system ($\tau_\Gamma$). We simplify our treatment by assuming that mechanical evolution ($\tau_{\mathrm{m},\Gamma}$) takes place on a timescale much longer than $\tau_\mathrm{r}$.}%
 \label{fig:Timescales}%
\end{figure}%
We claim that this process \emph{transfers the superposition from the particle state to the joint motional state of the mirrors}. Indeed, let us consider the timescales of the problem (cf.\ \fref{fig:Timescales}). The collision time $\tau_\mathrm{c}$ can be calculated to be of the order of $10^{-12}$\,s. For typical oscillator frequencies and cavity decay rates, therefore, $\tau_\mathrm{c}$ is effectively zero, even when $N\lesssim100$ particles are involved. The `information' imparted upon collision is stored in the form of phonons in the mirror structure. It is the decoherence timescale $\tau_\mathrm{d}$ of the phonon modes that sets the limit for the time $\tau_\mathrm{r}$ over which the optical read-out of the motion must take place: Over times of the order of $\tau_\mathrm{d}$, the system will lose its coherence through coupling to the mirror supports. The mismatch of mass between each mirror and the particles means that the transfer of momentum is very inefficient, with the vast majority of the energy going to exciting motional modes of the mirrors that carry no net forward momentum at time $t=0$. We assume that the particles impinge upon the mirrors in a balanced fashion around the centre, that they do not puncture the mirrors, and that there is a single high-quality low-frequency mode to which essentially all of the momentum is imparted. To simplify our description, the frequency of oscillation of this mode is taken to be such that $\tau_\mathrm{m}\gg\tau_\mathrm{r}$, with $\tau_\mathrm{m}$ the mechanical characteristic timescale. For a high-quality mode, the mechanical decay time $\tau_\Gamma$ is several orders of magnitude larger than $\tau_\mathrm{m}$, and we will therefore not be concerned with it.

\par
The state of the particles before collision ($t=0^-$) is described by the pure state
\begin{equation}
\ket{\psi}_\mathrm{part}={\binom{2N}{N}}^{\!\!-1/2}~\sum_{r=0}^N\binom{N}{r}\ket{N-r,r}_\mathrm{part}\,,
\end{equation}
whereas the motional state of each mirror is taken to be a thermal state with an average number of phonons $\bar{n}$, yielding the joint state $\rho_\mathrm{mech}(0^-)$. The collision process itself is modeled by a non-unitary operator that acts to transfer the momentum of the particles to the mirrors:
\begin{equation}
\label{eq:Collision}
\hat{C}\ket{n_1,n_2}_\mathrm{part}\to\hat{D}_1(\gamma n_1)\hat{D}_2(\gamma n_2)\,,
\end{equation}
where $\hat{D}_j(\alpha)=\exp[\sqrt2i(\im{\alpha}\hat{x}_j-\re{\alpha}\hat{p}_j)]$ is the displacement operator (of amplitude $\alpha\in\mathbb{C}$) for mechanical mode $j=1,2$, and $\hat{x}_j$ and $\hat{p}_j$ are its two quadrature operators. The factor $\gamma$ acts as a coupling strength parameter; we shall show below how $\gamma\sim1$ can be achieved in a realistic setup. The non-unitary nature of the transformation effectively encompassed by \eref{eq:Collision} can be understood in light of the fact that much of the energy goes to excite other mechanical modes that do not interact with the light field and are factored out of the description for $t\ll\tau_\mathrm{d}$. This is the reason for the quantum correlations that we will see occur at the mechanical level, despite the fact that the particle state is decomposed in terms of an orthogonal basis. The joint mechanical state of the mirrors just after the collision is therefore given by $\rho_\mathrm{mech}(0^+)=\tr_\mathrm{part}\bigl[{\hat{C}\,\rho(0^-)\,\hat{C}^\dagger}\bigr]/\tr\bigl[\hat{C}\,\rho(0^-)\,\hat{C}^\dagger\bigr]$, where the trace is taken over the particle degrees of freedom, and $\rho(0^-)=\ket{\psi}\bra{\psi}_\mathrm{part}\otimes\rho_\mathrm{mech}(0^-)$.\\
We reiterate that \eref{eq:Collision} is a valid description only if $\tau_\mathrm{r}\ll\tau_\mathrm{d}$. Over timescales of the order of $\tau_\mathrm{d}$, the system will decohere: `which way' information, containing the whereabouts of the ions that impinged on the mirrors, leaks out over this timescale through the supports of the mirrors. For times $\ll\tau_\mathrm{d}$, however, the ions cannot be localised on one mirror or the other, since this localisation is merely an effect of the decoherence brought about by the leaking out of this information into the environment. Bearing in mind that the collision process scrambles the ions' internal states, \eref{eq:Collision} incorporates the idea that the post-collision state of the ions is independent of $n_1$ and $n_2$.

\par
\begin{figure*}[th]%
\subfigure[\ (a)]{%
\includegraphics[width=0.5\figurewidth]{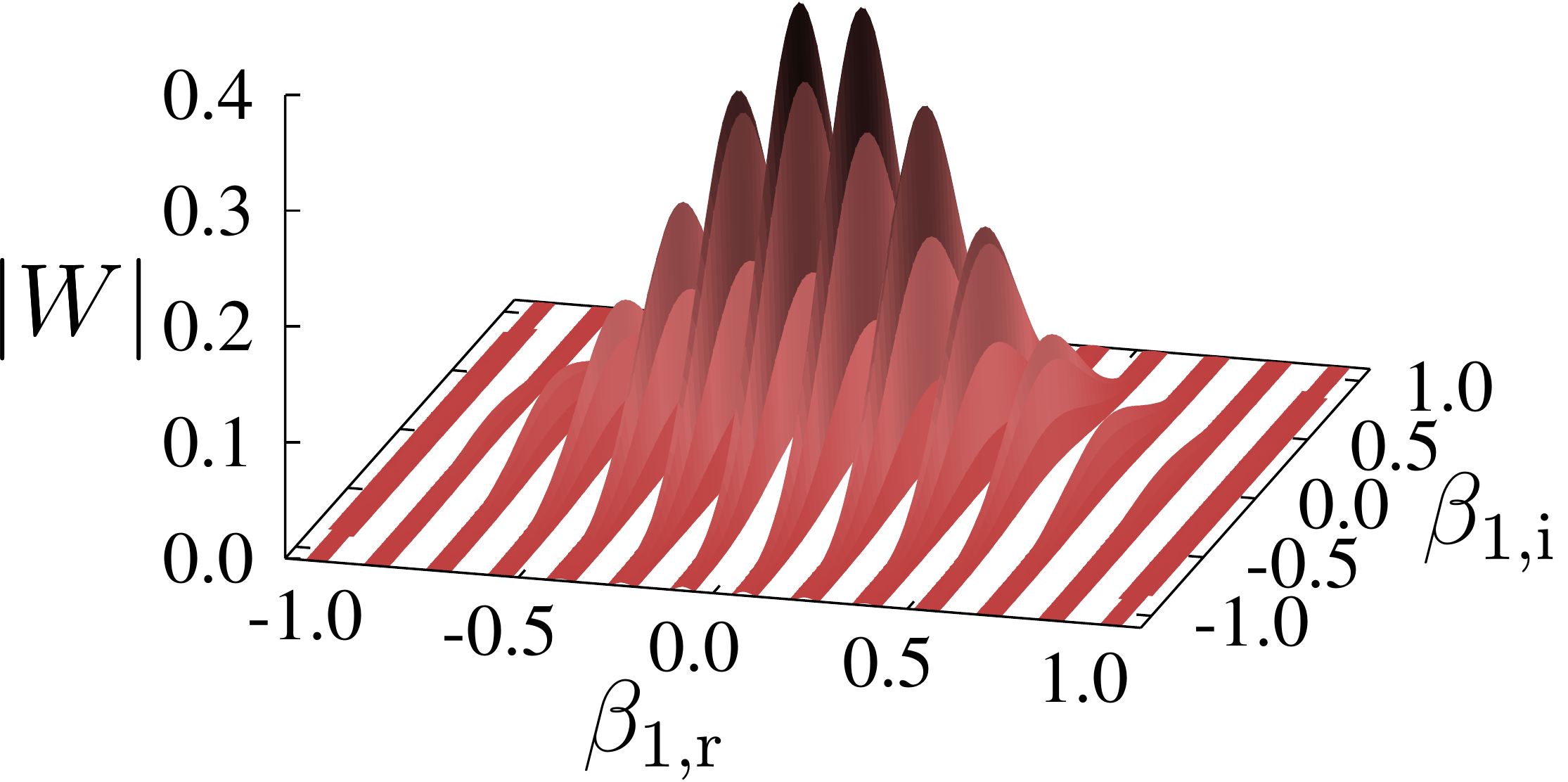}
}%
\qquad%
\subfigure[\quad\quad(b)]{%
\includegraphics[width=0.49\figurewidth]{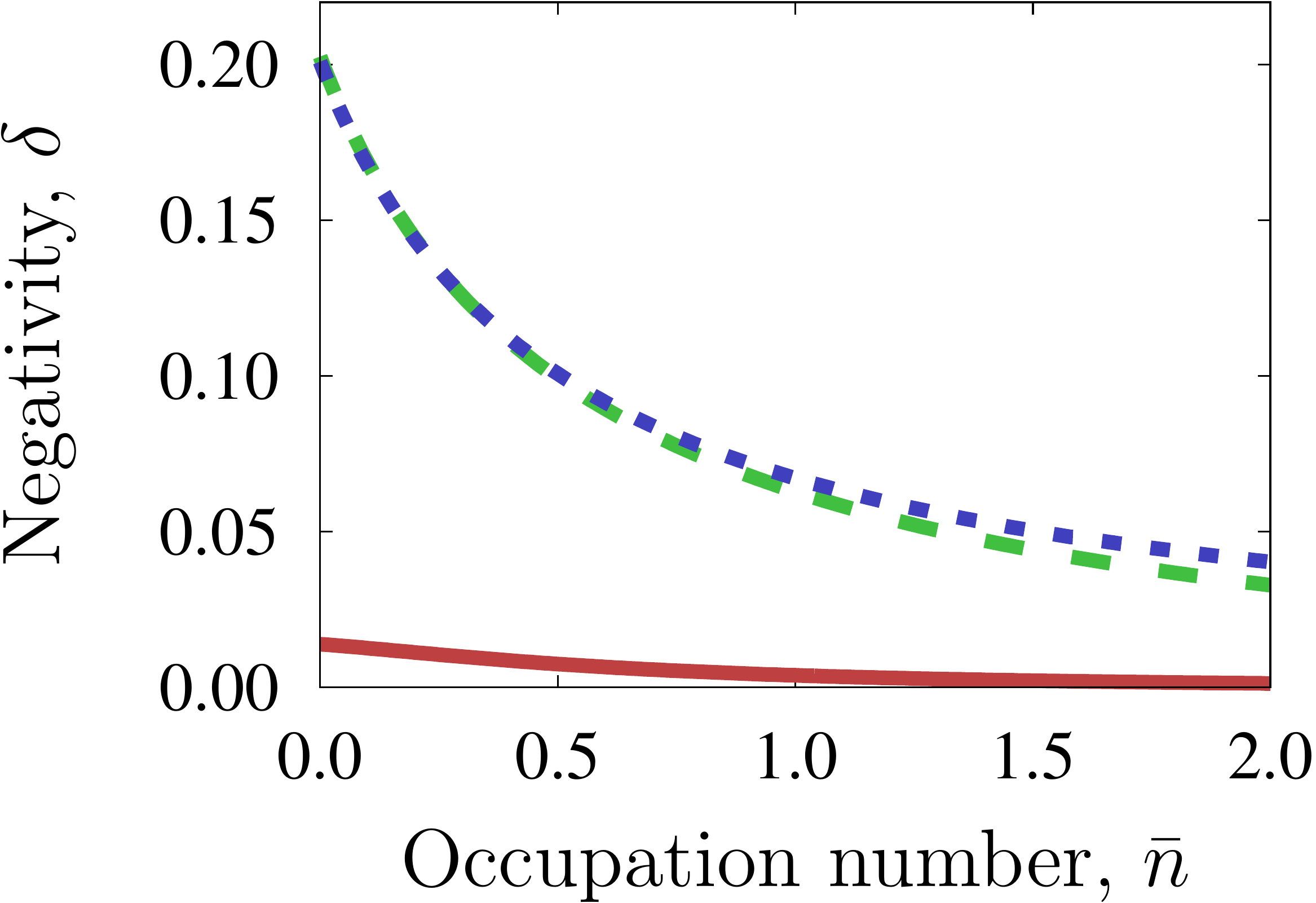}%
}%
\qquad%
\subfigure[\quad\quad(c)]{%
\includegraphics[width=0.49\figurewidth]{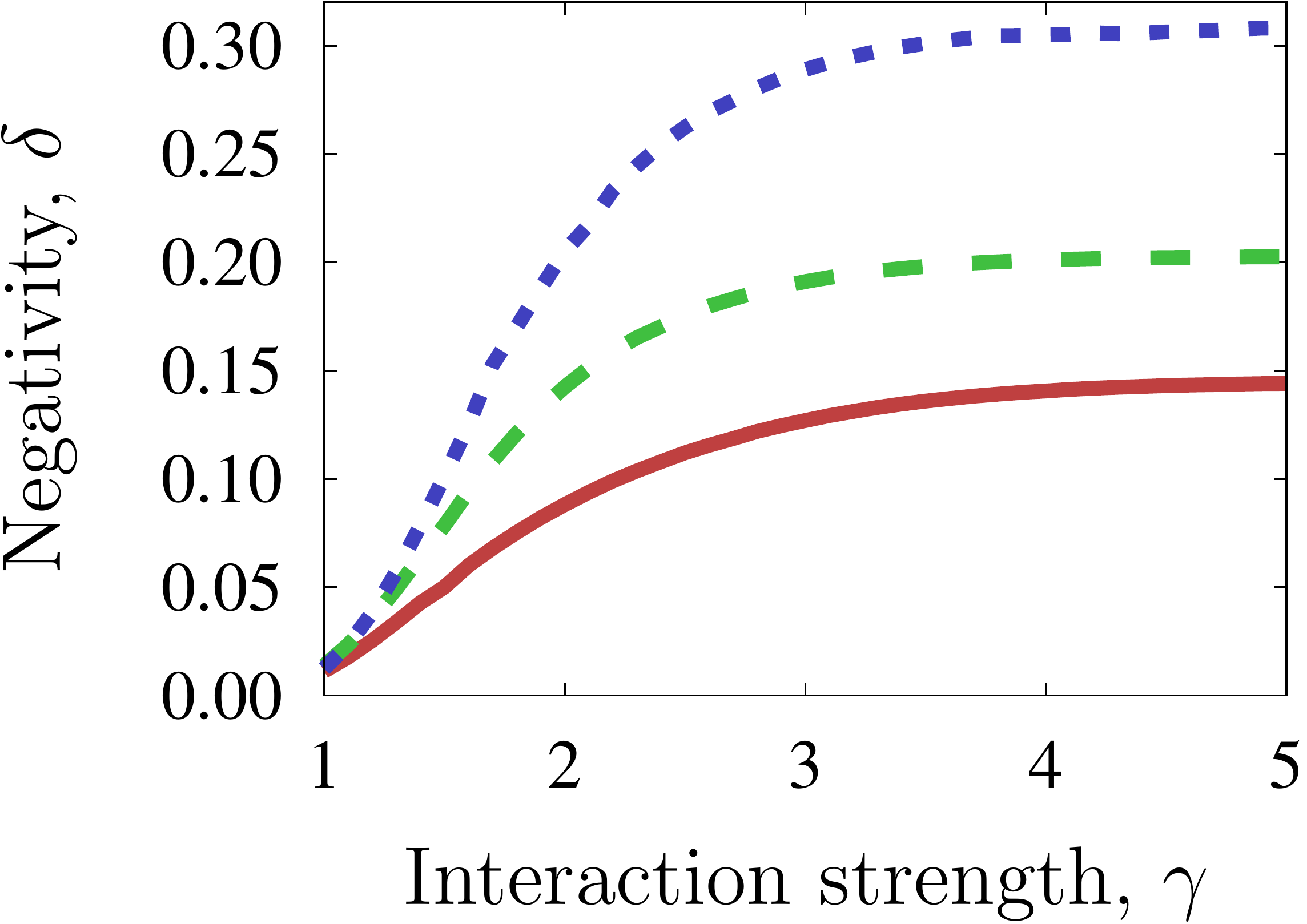}%
}%
\qquad%
\subfigure[\quad\quad(d)]{%
\includegraphics[width=0.49\figurewidth]{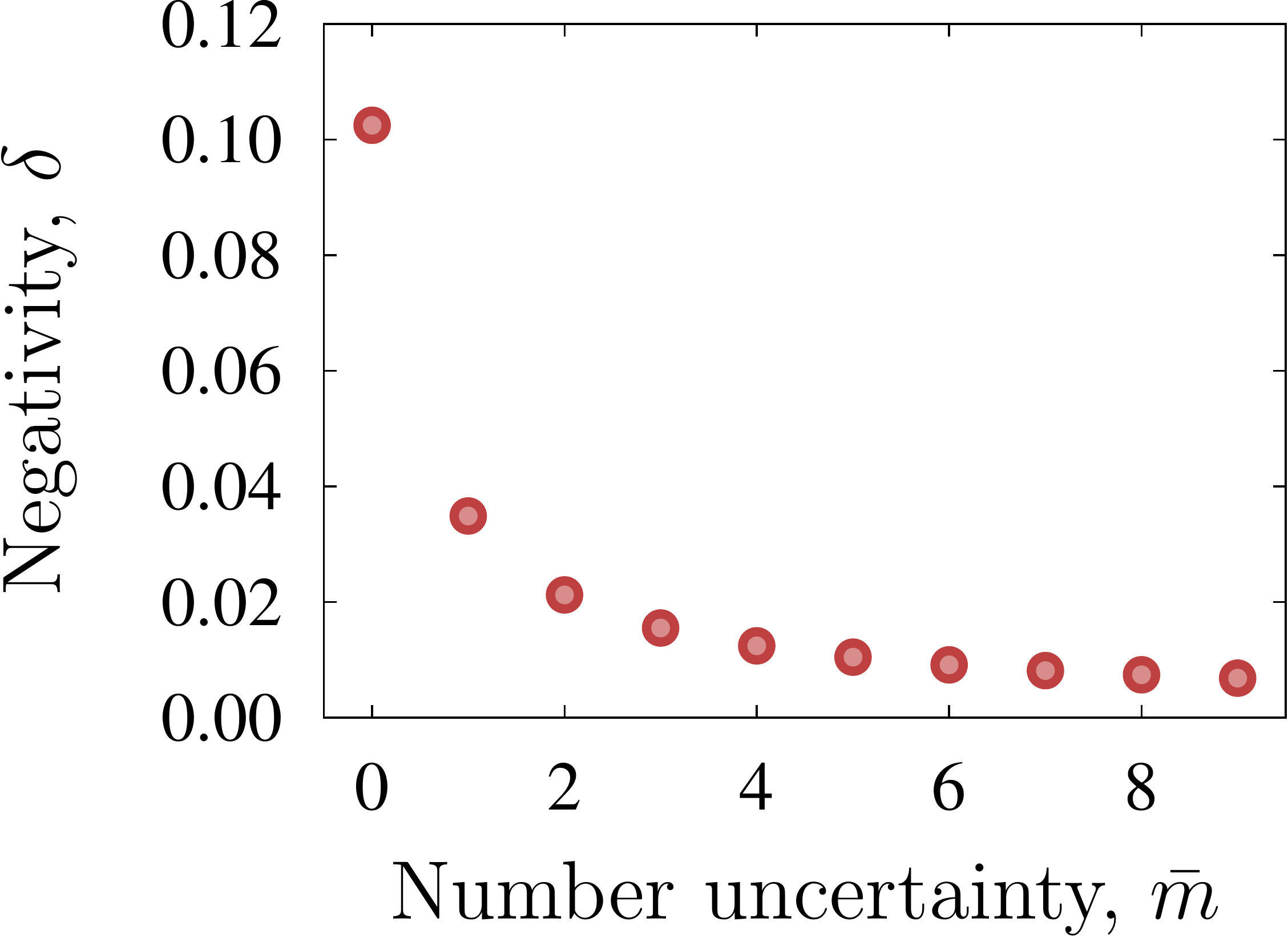}%
}%
 \caption{\textbf{Behaviour of the Wigner distribution negativity.} (a)~Modulus of the negative part of the mechanical Wigner distribution at the phase-space point $\beta_2=-\beta_1$ with $\beta_j=\beta_{j,\mathrm{r}}+i\beta_{j,\mathrm{i}}$ ($j=1,2$) for $\bar{n}=0$, $N=5$ and $\gamma=10$. (b)~Negativity versus $\bar{n}$ for $N=1$ and $\gamma=1$ (solid red curve), $5$ (dashed green), and $10$ (dotted blue). (c)~Negativity against $\gamma$ at $\bar{n}=0$ for $N=1$ (solid red curve), $2$ (dashed green), and $4$ (dotted blue). (d)~Negativity achieved, for $\gamma=5$, by a matter-wave beam with nominal number of particles $N=7$ and subjected to thermal fluctuations of size quantified by $\bar{m}$ (see text).}%
 \label{fig:Negativity}%
\end{figure*}

Recent experiments~\cite{Vanner2012b} have demonstrated the first steps towards tomographic readout of mechanical states. Such measurements give access to the Wigner--Weyl quasiprobability distribution $W(\beta_1,\beta_2)$ corresponding to $\rho_\mathrm{mech}(0^+)$, where $\beta_{1,2}$ are the two phase-space coordinates for the mechanical systems. One crucial quantity we aim to calculate is the \emph{negativity}~\cite{Kenfack2004} $\delta=\iint\rmd^2\beta_1\rmd^2\beta_2\,\abs{W(\beta_1,\beta_2)}-1$. The Wigner distribution of any convex sum of displaced (coherent) states is non-negative. A value $\delta>0$ therefore witnesses non-classical correlations. An evaluation of $W(\beta_1,\beta_2)$ for appropriate $\beta_{1,2}$, as well as the corresponding values of $\delta$ is shown in \fref{fig:Negativity}(a)--(c). We see that $\delta$ is indeed nonzero and, as expected, the negative volume of the Wigner distribution grows with $\gamma$ and decreases with the thermal character of the mechanical state. This demonstrates that strong non-classicality can be induced in the state of the mechanical system by the quantum coherence enforced in the matter-wave part of our setting. Pre-cooling of the mechanical system, using optomechanics or traditional refrigeration, is necessary to mitigate the effects of a large $\bar{n}$. Moreover, such non-classical character appears to survive in the presence of imperfections in the matter-wave resource state, e.g., due to a thermally fluctuating number of particles per arm of the interferometer. That is, we can consider a total number of particles fluctuating around a nominal value $N$ according to a thermal distribution of mean $\bar{m}$ and study the behaviour of $\delta$ against such uncertainty. \fref{fig:Negativity}(d) shows that, albeit depleted, $\delta$ persists to significant fluctuations of $N$.\par
\begin{figure}[bh]%
\includegraphics[width=\figurewidth]{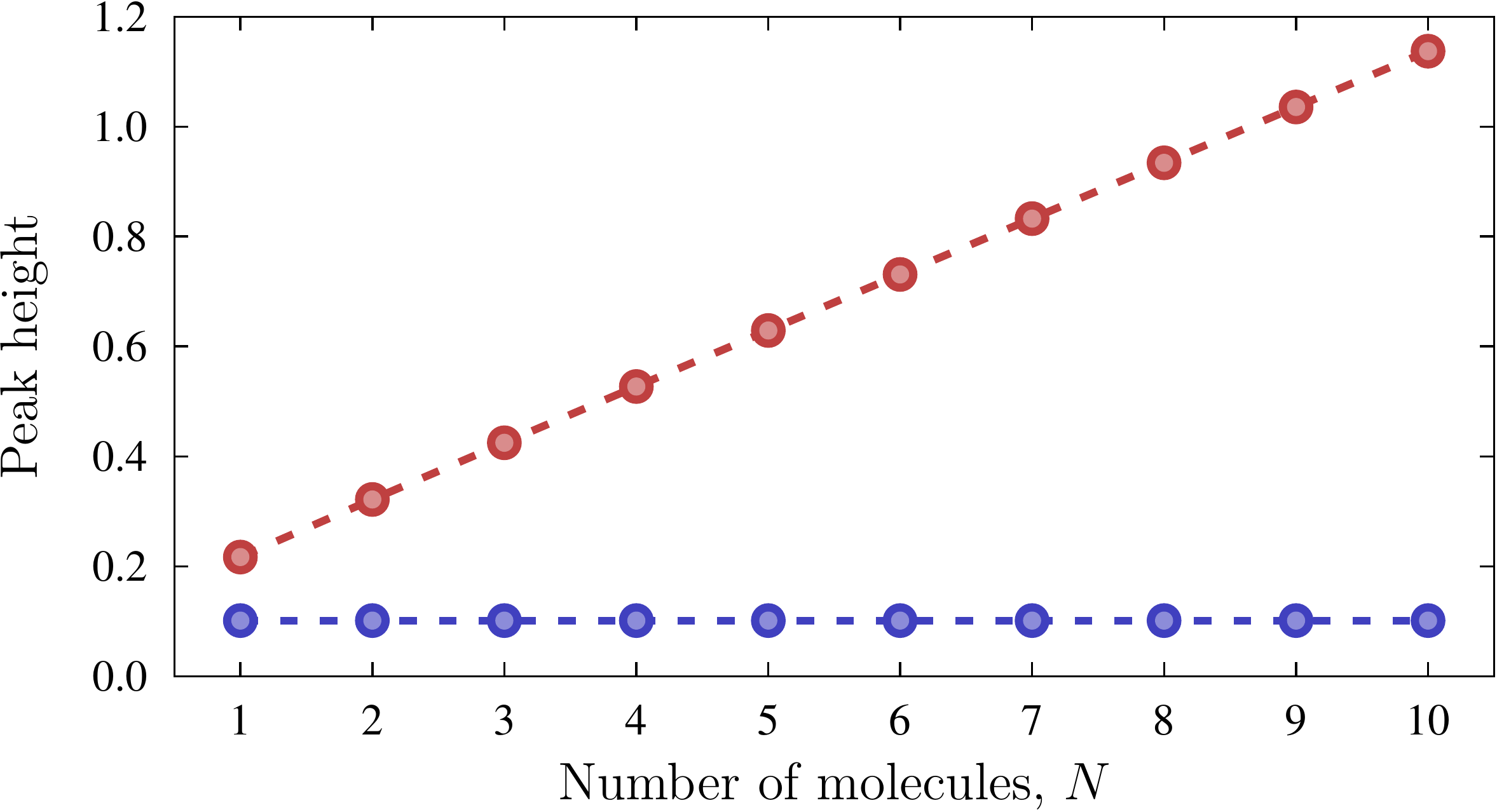}%
 \caption{\textbf{Behaviour of resonance with number of particles.} We plot the height of the resonance peak $P(x_\theta,y_\phi)$ as a function of the number of particles $N$. The height of the peak does not vary in the incoherent case (blue points), but increases linearly in the coherent case (red points). We took $\bar{n}=0,\gamma=1$.}%
 \label{fig:PeakHeights}%
\end{figure}

\begin{figure}[bh]%
\includegraphics[width=\figurewidth]{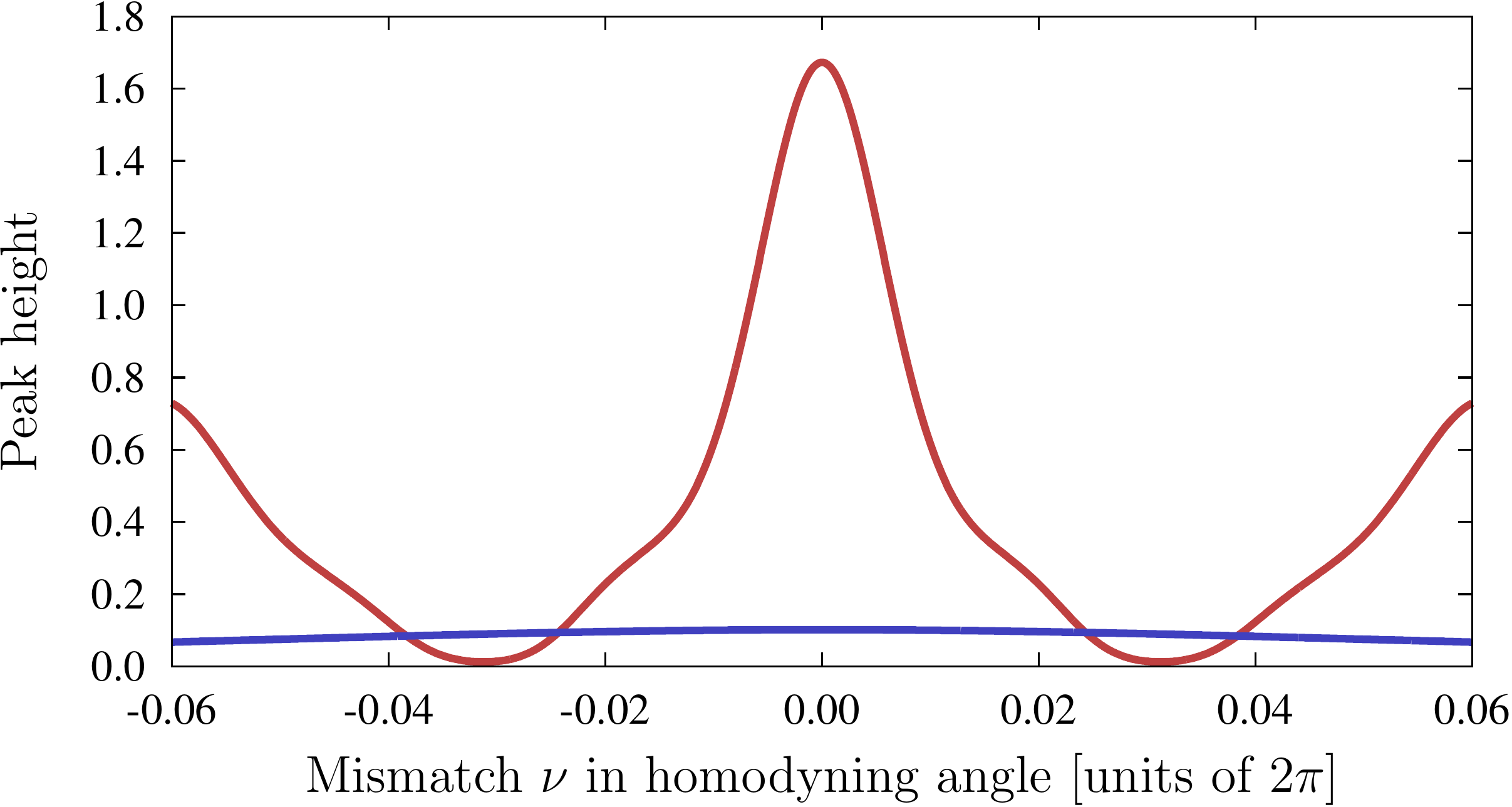}%
 \caption{\textbf{Effect of homodyning angle on resonance peak.} Here we show the height of the resonance peak in $P(x_\theta,y_\phi)$ as a function of the offset in homodyning angle $\nu$. The peak height varies monotonically with $\nu$ in the incoherent case (blue curves), but may exhibit revivals in the coherent case (red). We took $N=5,\bar{n}=0,\gamma=5$.}%
 \label{fig:PeakVariation}%
\end{figure}

We shall now discuss two further signatures that bear witness to the preservation of correlations in the post-collision joint mechanical state. We superimpose the readout fields on a balanced beamsplitter and examine the value of $P(x_\theta,y_\phi)=\abs{\bra{x_\theta,y_\phi}\hat{B}\,\hat{\rho}_\mathrm{mech}(0^+)\,\hat{B}^\dagger\ket{x_\theta,y_\phi}}^2$ where $\hat{B}=\exp[i\frac{\pi}{4}(\hat{x}_1\hat{x}_2+\hat{p}_1\hat{p}_2)]$ is the beamsplitter operator and $\ket{x_\theta,y_\phi}$ is the joint mechanical state in the position representation. Interference effects in $P(x_\theta,y_\phi)$, due to coherences in the density matrix, arise such that the terms making it up may add constructively or destructively. For certain measurement settings, the behaviour of $P(x_\theta,y_\phi)$ around its local maximum carries two signatures of these coherences. First, the value of $P(x_\theta,y_\phi)$ at the maximum itself is independent of $N$ for a classical ensemble where $\rho_\mathrm{part}(0^-)$ is diagonal. The contribution of the off-diagonal terms, however, is such that the value of the function increases approximately linearly with $N$ for a coherent superposition (cf.\ \fref{fig:PeakHeights}).\\
Second, the transition between constructive and destructive interference when the homodyne angles are each incremented by a small value causes the peak in $P(x_\theta,y_\phi)$ to oscillate in amplitude. A diagonal $\rho_\mathrm{part}(0^-)$ yields quantitatively different behaviour, as shown in \fref{fig:PeakVariation}. Finally, we note that the location of the maximum depends only on $\gamma N$, which can be controlled by changing the acceleration potential.

\noindent\textbf{Light--matter-wave interface.} The use of a beamsplitter to mix the output signals from the optomechanical cavities opens the door to a coherent coupling between the mechanical and the optical systems. This can be achieved through the use of polarisation optics in conjunction with the beamsplitter, whereby the coherence of the state is maintained upon readout by the optical field, as suggested by the read-out optics shown in \fref{fig:System}. Together with the coherent coupling between the matter wave and the mechanical state, this provides for the possibility of a coherent (albeit indirect) coupling between matter waves and light. The presented system thus forms the basis of a coherent interface for integrating matter waves into quantum networks~\cite{Kimble2008}, a possibility that we will explore in the future.

\noindent\textbf{Reflective coupling.} Thus far, we have described an interface based on the adsorption of particles on the mirror surface. Whilst technically simpler than a reflective interface, this differs substantially from the usual OM paradigm of photons reflecting off a mirror surface. Electrostatic ion mirrors can be constructed~\cite{McLuckey2001}, although their integration into an OM setup as sketched in \fref{fig:System} would be challenging. A distinct advantage of operating with a reflective, rather than adsorptive, coupling is that one can use the system in a time-reversed fashion, where the optomechanical elements \emph{imprint} information coherently onto the matter wave. Arbitrary states of light can be coherently transferred to the mechanical system through the `optomechanical beamsplitter' interaction~\cite{Aspelmeyer2010}. In a somewhat analogous manner we then envisage a further state-transfer step, where the mechanical state is transferred to an impinging matter-wave field upon its reflection. This would open the door to a recombination of the two paths of the matter-wave interferometer after one, or perhaps both, have been coherently manipulated. The possibility of a coherent bidirectional coupling between optics and matter waves therefore raises a number of interesting possibilities, including the indirect interaction of different matter waves, perhaps belonging to entirely different species and separated widely in space or time, through the intermediate storage of coherences in the optomechanical systems.

\noindent\textbf{Experimental viability.} The strength of the effects we discussed above is determined mostly by the single parameter $\gamma$, which quantifies the displacement in phase space incurred by the mirror upon adsorption of a single particle. Consider a particle of de Broglie wavelength $\lambda_\mathrm{dB}$ and a mobile mirror whose zero-point fluctuations have an extent $x_\mathrm{zpt}$. Modeling the collision process as a $\delta$-function in time yields $\gamma=\sqrt{2}\pi\times x_\mathrm{zpt}/\lambda_\mathrm{dB}$. To estimate the size of $\gamma$, we shall suppose that our mobile mirrors have oscillation frequency $\omega_\mathrm{m}=2\pi\times150$\,kHz and mass $M=1.4\times10^{-13}$\,kg, yielding $x_\mathrm{zpt}\approx29$\,fm. A single helium atom has a mass $m\approx6.6\times10^{-27}$\,kg and, can be accelerated through a potential $V=10$\,kV to yield $\lambda_\mathrm{dB}\approx144$\,fm. The high speed of the ions will be helpful to maintain the coherence of the particle during acceleration and manipulation before it hits the mirror, as per the discussion of decoherence effects in Ref.~\onlinecite{Hasselbach2010}. Putting these numbers together gives $\gamma\approx0.9$. The mirror oscillation induced by a single ion collision results in a shift of the resonance frequency of the cavity by a maximum amount $\Delta\omega=2\pi\times\omega_\mathrm{c}x_\mathrm{zpt}^2/(L_\mathrm{c}\lambda_\mathrm{dB})$, which, for a cavity length $L_\mathrm{c}=400\lambda_\mathrm{c}$ and central frequency $\omega_\mathrm{c}=2\pi c/\lambda_\mathrm{c}$ (corresponding to $\lambda_\mathrm{c}=1064$\,nm), gives $\Delta\omega\approx2\pi\times24$\,kHz. This value must be compared to the linewidth of the cavity which, for a finesse of $7000$, would be $\kappa_\mathrm{c}\approx2\pi\times25$\,MHz. The measurement of the effect of a single particle is thus very close to current technological possibilities.

\section*{Discussion}
We have presented a system that combines optomechanics and matter-wave interferometry to provide a coherent interface between optics and matter waves. Because of the universality of optical interfaces, our system expands the range of applicability of matter-wave interferometry with massive molecules beyond fundamental tests of quantum mechanics and metrology. It opens the door to coherently embedding matter waves into quantum networks consisting of optical or hybrid constituents, and potentially enables the bidirectional transfer of quantum information between light and matter waves. As such, our work opens new possibilities for studies spanning from hybrid quantum communication to tests of fundamental quantum mechanics in setups that put together the advantages inherent in various realisations of quantum interferometers.

{\footnotesize
\section*{Methods}
In this section we outline the main steps needed to reproduce our calculations. Further details and the intermediate steps are shown in the supplementary information.

\noindent\textbf{Modeling a stream of particles.} The action of the biprisms in the setup sketched in the main text is similar to that of a balanced beamsplitter on an optical field. Thus, after interacting with the biprism, a single incoming particle can be described through the superposition state
\begin{equation}
\frac{1}{\sqrt{2}}\bigl(\ket{0,1}_\mathrm{part}+\ket{1,0}_\mathrm{part}\bigr)=\sum_{r=0}^1\binom{1}{r}\ket{1-r,r}_\mathrm{part}\,.
\end{equation}
The collision process destroys the ions' internal states, so the pre-collision state needs only to keep track of the number of ions in each arm of the interferometer, rather than their positions or internal states. With this in mind, in the supplementary information we show that the state of a stream of $N$ particles after traversing the biprism can be written as
\begin{equation}
\label{eq:PartPostBiprism}
\ket{\psi}_\mathrm{part}\equiv\sum_{r=0}^N\binom{N}{r}\ket{N-r,r}_\mathrm{part}\,,
\end{equation}
assuming that the total time separating the arrival of the first and last particles is very small compared to all the other timescales of the problem. Under these conditions, multiple `bursts' can be treated \emph{identically} to a single burst with the same total number of particles.

\noindent\textbf{Post-collision density matrix.} We begin with the particle state $\ket{\psi}_\mathrm{part}$, and the corresponding density matrix $\rho_\mathrm{part}(0^-)=\ket{\psi}\bra{\psi}_\mathrm{part}$ corresponding to this state can be written. The initial density matrix for the mechanical systems is taken to be identical, i.e., a thermal state with average phonon occupation $\bar{n}$:
\begin{equation*}
\rho_\mathrm{mech}(0^-)=\frac{1}{\pi^2\bar{n}^2}\iint{\rmd^2\beta_1\rmd^2\beta_2}\,e^{-\tfrac{\abs{\beta_1}^2+\abs{\beta_2}^2}{\bar{n}}}\ket{\beta_1,\beta_2}\,{\bra{\beta_1,\beta_2}_\mathrm{mech}}\,,
\end{equation*}
expressed in terms of joint coherent states. Subsequently we obtain the density matrix for the total system just before the collision occurs: $\rho(0^-)=\rho(0^-)_\mathrm{part}\otimes\rho(0^-)_\mathrm{mech}$. We model the collision process as described in the main text, yielding the normalised post-collision joint mechanical density matrix $\rho_\mathrm{mech}(0^+)$. Several calculations yield strikingly different results between the coherent and incoherent (where the matrix is decohered in the Fock basis) situations.

\noindent\textbf{Joint homodyne detection.} Let us suppose we can read out any quadrature of the mechanical oscillator state through the cavity field. We would like to calculate the possibility of obtaining a joint measurement with amplitudes $x$ and $y$ and phases $\theta$ and $\phi$. Instead of reading out the individual cavity fields, however, we interfere them on a beamsplitter prior to homodyning. The beamsplitter operator, which we represent by $\hat{B}$ and define in the main text, effectively acts on the mechanical fields in the coherent representation as
\begin{equation}
\hat{B}\ket{\beta_1,\beta_2}=\ket{(\beta_1+i\beta_2)/\sqrt{2},(i\beta_1+\beta_2)/\sqrt{2}}\,.
\end{equation}
After operation by the beamsplitter, the state is projected onto the quadratures $x_\theta$ and $y_\phi$, and the absolute-squared value of the projection measured, $P(x_\theta,y_\phi)$, as defined in the main text.

\noindent\emph{Behaviour of resonance with number of particles.} We choose $\sin\theta=-\cos\theta=\pm\tfrac{1}{\sqrt{2}}$ and monitor the coincidences for which $x=y$. In this case, when $\phi=-\theta-\tfrac{\pi}{2}$. After some algebra, we get to 
\begin{equation}
P(x_\theta,y_\phi)=\begin{cases}
\frac{1}{\pi^2(2\bar{n}+1)^2}&\\
\frac{2^{4N}}{\pi^2(2\bar{n}+1)^2}\Bigl[\sum_{r=-N}^N\binom{2N}{N+r}e^{-(2\bar{n}+1)\gamma^2r^2}\Bigr]^{-2}\,.
\end{cases}
\end{equation}
The former of these equations (the incoherent case) is \emph{independent of the number of particles or the strength of the interaction}, whereas the latter (coherent) is approximately linear in $N$.

\noindent\emph{Effect of homodyning angle on the resonance peak.} With the choice of parameters analysed above, $P(x_\theta,y_\phi)$ has Gaussian peaks for both coherent and the incoherent cases. Let us now set $\theta=\theta_0+\nu$ and $\phi=-\theta_0-\tfrac{\pi}{2}+\nu$. It can be shown that, to lowest order in $\nu$, there is again a qualitative difference between the two cases: The peak in the incoherent case is constant with respect to small variations in $\nu$, whereas the coherent case exhibits characteristic variations with such parameter.

\noindent\textbf{Momentum imparted by a colliding particle.} Let us work in a classical picture to estimate the effect of the adsorption of a single particle on a mirror. The change in (dimensionless) momentum of the mirror, $\Delta p$, due to the instantaneous adsorption of a particle travelling with momentum $p_\mathrm{part}$, is given by $M\omega_\mathrm{m}x_\mathrm{zpt}\Delta p=-p_\mathrm{part}$. Here $M$ is the mass of the mirror and $\omega_\mathrm{m}$ its oscillation frequency. We can write this as $\Delta p=-G$ with 
\begin{equation}
G=2\pi\times\frac{x_\mathrm{zpt}}{\lambda_\mathrm{dB}}=\frac{p_\mathrm{part}}{\sqrt{\hbar M\omega_\mathrm{m}}},
\end{equation}
where $\lambda_\mathrm{dB}=h/p_\mathrm{part}$ is the de Broglie wavelength of the matter wave. The extent of the zero-point fluctuations of the mirror motion is $x_\mathrm{zpt}=\sqrt{\hbar/(M\omega_\mathrm{m})}$. We divide $G$ by a factor of $\sqrt{2}$, in accordance with the definition of $\hat{D}_{1,2}$, to obtain the parameter $\gamma$ used in the paper.

\noindent\textbf{Frequency shift due to a single collision.} We work in a classical picture to approximate the frequency shift obtained in a resonant cavity as the result of the adsorption of a single particle. The momentum imparted onto the mirror results in a maximal displacement $\Delta L$ such that $M\omega_\mathrm{m}\Delta L=p_\mathrm{part}\Rightarrow\Delta L={p_\mathrm{part}}/(M\omega_\mathrm{m})$.
For a resonant cavity, we have $L_\mathrm{c}=\tfrac{n}{2}\lambda_\mathrm{c}$, where $n\in\mathbb{N}$. Thus,
\begin{equation}
\Delta L=\tfrac{n}{2}\Delta\lambda\Rightarrow\frac{\Delta L}{L_\mathrm{c}}=\frac{\Delta\lambda}{\lambda_\mathrm{c}}\,.
\end{equation}
We can therefore calculate the maximum frequency modulation
\begin{equation}
\omega_\mathrm{c}=\frac{2\pi c}{\lambda_\mathrm{c}}\Rightarrow\Delta\omega=\frac{\omega_\mathrm{c}}{L_\mathrm{c}}\lvert\Delta L\rvert=\frac{\omega_\mathrm{c}p_\mathrm{part}}{L_\mathrm{c}\omega_\mathrm{m}M}=2\pi\times\frac{\omega_\mathrm{c}x_\mathrm{zpt}^2}{L_\mathrm{c}\lambda_\mathrm{dB}}\,.
\end{equation}
}

\bibliographystyle{naturemag}

{\footnotesize
\ \\
\noindent{\small\textbf{Acknowledgements}}\\
We are grateful to Riccardo Sapienza and Giuseppe Vallone for valuable suggestions. We acknowledge funding from the UK EPSRC through a Career Acceleration Fellowship and a grant under the ``New Directions for EPSRC Research Leaders'' initiative (EP/G004579/1), the J.\ Templeton Foundation through the Foundational Questions Institute (FQXi) and the Grant 43467, the South-English Physics network (SEPNet), the Alexander von Humboldt Stiftung, and the Royal Commission for the Exhibition of 1851.

\ \\
\noindent{\small\textbf{Additional information}}\\
\textbf{Supplementary information} accompanies this paper at http://www.nature.com/scientificreports\\
\ \\
\textbf{Author contributions:} AX, HU, and MP jointly developed the concept, interpreted the results, and wrote the manuscript. AX and MP generated the data for Figs.\ 3--5.\\
\ \\
\textbf{Competing financial interests:} The Authors declare no competing financial interests.\\
\ \\
PACS numbers: 37.25.+k 03.75.Dg 42.50.Wk
}

\appendix\begin{widetext}
\section{Modeling a stream of particles}
The action of the biprisms in the setup sketched in the main text is similar to that of a balanced beamsplitter on an optical field. Thus, after interacting with the biprism, a single incoming particle can be described through the superposition state
\begin{equation}
\frac{1}{\sqrt{2}}\bigl(\ket{0,1}_\mathrm{part}+\ket{1,0}_\mathrm{part}\bigr)=\sum_{r=0}^1\binom{1}{r}\ket{1-r,r}_\mathrm{part}\,.
\end{equation}
Here we will show that the state of a stream of $N$ particles after traversing the biprism can be written as
\begin{equation}
\label{eq:PartPostBiprism}
\ket{\psi}_\mathrm{part}\equiv\sum_{r=0}^N\binom{N}{r}\ket{N-r,r}_\mathrm{part}\,,
\end{equation}
assuming that the total time separating the arrival of the first and last particles is very small compared to all the other timescales of the problem (as per the discussion in the main text). Suppose we fire two bursts of $N_1$ and $N_2$ particles at the two mirrors, with the time separation between the two bursts being negligible compared to $\tau_\mathrm{m}$. Then, the particle state after the biprism reads
\begin{align}
\ket{\psi}_\mathrm{part}&=\sum_{r_1=0}^{N_1}\sum_{r_2=0}^{N_2}\binom{N_1}{r_1}\binom{N_2}{r_2}\ket{(N_1+N_2)-(r_1+r_2),(r_1+r_2)}_\mathrm{part}\nonumber\\
&=\sum_{r=0}^{N}\sum_{\substack{r^\prime=-N_2,\\(r\pm r^\prime)\text{\ even}}}^{N_1}\binom{N_1}{(r+r^\prime)/2}\binom{N_2}{(r-r^\prime)/2}\ket{N-r,r}_\mathrm{part}\,,
\end{align}
where $N=N_1+N_2$. Next, we use the fact that
\begin{equation}
\binom{n}{k}=0
\end{equation}
whenever $k<0$ or $k>n$, as follows from the definition of the Euler Gamma function, which diverges for non-positive integer values. This fact allows us to rewrite the sum above as
\begin{equation}
\sum_{\substack{r^\prime=-N_2,\\(r\pm r^\prime)\text{\ even}}}^{N_1}\cdots\quad\equiv\sum_{\substack{r^\prime=-r,\\(r\pm r^\prime)\text{\ even}}}^{r}\cdots\,,
\end{equation}
since $r^\prime>N_1$ implies that the only terms that might contribute require $r\geq r^\prime>N_1$, such that $(r+r^\prime)/2>N_1$; i.e., every such term is zero. Similarly, terms with $r^\prime<-N_2$ do not contribute. It does not change anything, therefore, to extend the lower (upper) limit of the sum to $-r$ ($r$) when $r>N_2$ ($r>N_1$). Moreover, the requirement for the bottom factor of the binomials to be non-negative requires $\abs{r^\prime}\leq r$. Therefore, the lower (upper) limits of the sum may be changed to $-r$ ($r$) in any case. Let us now identify $n\equiv N_1$, $m\equiv N_2$, $k\equiv(r+r^\prime)/2$, and $p\equiv r$, all of which are non-negative integers. From Ref.~\cite[\S0.156 Eq.~(1)]{Gradshteyn1994} we know that
\begin{equation}
\sum_{k=0}^p\binom{n}{k}\binom{m}{p-k}\equiv\binom{n+m}{p}\,,
\end{equation}
that is,
\begin{equation}
\sum_{\frac{r+r^\prime}{2}=0}^r\binom{N_1}{(r+r^\prime)/2}\binom{N_1}{(r-r^\prime)/2}\equiv\binom{N_1+N_2}{r}\equiv\binom{N}{r}\,,
\end{equation}
where the sum runs over the values of $r^\prime$ for which $r+r^\prime$ is even, since $k$ must be an integer. Moreover, $0\leq k\leq p$, which is equivalent to $-r\leq r^\prime\leq r$. Thus, we can rewrite
\begin{equation}
\sum_{\frac{r+r^\prime}{2}=0}^r\cdots\quad\equiv\sum_{\substack{r^\prime=-r,\\(r\pm r^\prime)\text{\ even}}}^{r}\cdots\,,
\end{equation}
which gives \eref{eq:PartPostBiprism}. In other words, proceeding by induction, multiple `bursts' can be treated \emph{identically} to a single burst with the same total number of particles.

\section{Effect of collisions on mirror}
\begin{figure}[t]%
 \includegraphics[width=\figurewidth]{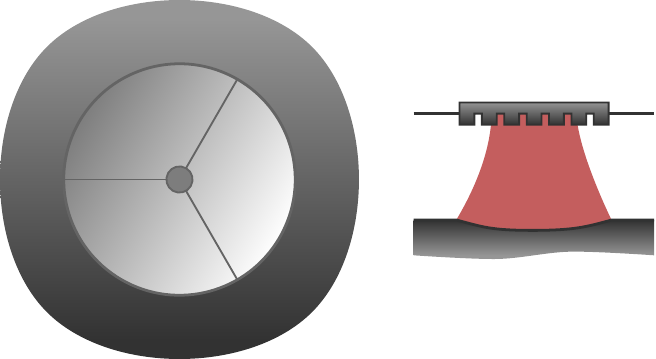}%
 \caption{Example mirror structure having properties that are compatible with the requirements of our thought experiment. The figure on the left shows a plan of the mirror, with the large (immobile) cavity mirror in the background. On the right, we show a section of the optical mode inside the microcavity; the mobile mirror may be patterned~\cite{Kemiktarak2012} in order to increase its reflectivity.}%
 \label{fig:Mirror}%
\end{figure}
The estimates in the main text resulted from calculations using structure shown in \fref{fig:Mirror}. In particular, we suppose that the central (reflective) part of the suspended mirror is circular with diameter $10$\,$\upmu$m and thickness $500$\,nm, and made from silicon nitride (yielding a mass of $1.3\times10^{-13}$\,kg for the mirror itself). The three mirror supports are assumed to be robust enough for nonlinear effects to be ignored. The fundamental motion of such a system, to a good approximation, involves the circular mirror itself undergoing centre-of-mass oscillations. A rudimentary finite element analysis of this structure shows that at the start of the harmonic cycle, only the fundamental mode carries a significant amount of net forward momentum. This leads us to postulate that upon collision, whereas only a small fraction of the energy of the particle is transferred to the fundamental mode essentially all of the momentum is. This argument relies on the mirror not being punctured by the particles; we follow Figs.~3 and~4 in Ref.~\cite{Miyagawa1984} as an approximation to our situation and determine that a thickness of $500$\,nm should be enough to stop the incident ions. The collision time stated in the main text is an approximated upper bound, determined as the time it would take for the particles to come to a complete stop over a distance of $500$\,nm. The rest of the energy imparted to the mirror through the collision process goes to excite higher-order modes, most of which are coupled to the optical fields significantly more weakly than the fundamental mode. Insofar as the optical readout process of the fundamental mechanical mode happens on a faster timescale than the decoherence timescale relevant to these mechanical modes, as explained in the main text, the excitation of these modes should not affect the results reported in the main text.

\section{Post-collision density matrix}
We begin recalling the form of the particle state $\ket{\psi}_\mathrm{part}={\binom{2N}{N}}^{\!\!-1/2}~\sum_{n=0}^N\binom{N}{n}\ket{N-n,n}_\mathrm{part}$. The density matrix $\ket{\psi}\bra{\psi}_\mathrm{part}$ corresponding to this state can be written as
\begin{equation}
\rho_\mathrm{part}(0^-)=\Biggl[\sum_{r=0}^N\binom{N}{r}^2\Biggr]^{-1}\sum_{r_1,r_2=0}^N\binom{N}{r_1}\binom{N}{r_2}\Phi(r_1-r_2)\ket{N-r_1,r_1}\bra{N-r_2,r_2}_\mathrm{part}\,,
\end{equation}
where the function $\Phi(r)$ acts as a control that changes $\rho_\mathrm{part}(0^-)$ from a coherent superposition [$\Phi(r)=1$] to an incoherent mixture [$\Phi(r)=\delta_{r,0}$, where $\delta_{i,j}$ is the Kronecker delta]. Combining this with the density matrix
\begin{equation*}
\rho_\mathrm{mech}(0^-)=\frac{1}{\pi^2\bar{n}^2}\iint{\rmd^2\beta_1\rmd^2\beta_2}\,e^{-\tfrac{\abs{\beta_1}^2+\abs{\beta_2}^2}{\bar{n}}}\ket{\beta_1,\beta_2}\,{\bra{\beta_1,\beta_2}_\mathrm{mech}}\,,
\end{equation*}
expressed in terms of joint coherent states, and where the mirrors are taken to be in thermal states with average phonon occupation $\bar{n}$ (assumed to be the same for both mirrors; whilst this assumption is not crucial for our proposal, it allows us to simplify our formal treatment of the problem), we obtain the density matrix for the total system just before the collision occurs
\begin{multline}
\rho(0^-)=\rho(0^-)_\mathrm{part}\otimes\rho(0^-)_\mathrm{mech}=\frac{1}{\pi^2\bar{n}^2\sum_{r=0}^N\binom{N}{r}^2}\sum_{r_1,r_2=0}^N\binom{N}{r_1}\binom{N}{r_2}\Phi(r_1-r_2)\\
\times\iint\rmd^2\beta_1\,\rmd^2\beta_2\,e^{-(\abs{\beta_1}^2+\abs{\beta_2}^2)/\bar{n}}\ket{N-r_1,r_1}\ket{\beta_1,\beta_2}\bra{\beta_1,\beta_2}\bra{N-r_2,r_2}\,.
\end{multline}
We model the collision process as described in the main text, yielding the normalised post-collision joint mechanical density matrix
\begin{multline}
\rho_\mathrm{mech}(0^+)=\sum_{r_1,r_2=0}^N\binom{N}{r_1}\binom{N}{r_2}\frac{\Phi(r_1-r_2)}{\pi^2\bar{n}^2\mathcal{Q}}\iint\rmd^2\beta_1\,\rmd^2\beta_2\,e^{-\frac{\abs{\beta_1}^2+\abs{\beta_2}^2}{\bar{n}}}e^{-i\gamma(r_1-r_2)\re{\beta_1-\beta_2}}\\
\times\ket{\beta_1+i\gamma(N-r_1),\beta_2+i\gamma r_1}\bra{\beta_1+i\gamma(N-r_2),\beta_2+i\gamma r_2}\,,
\end{multline}
where $\mathcal{Q}=\sum_{r_1,r_2=0}^N\binom{N}{r_1}\binom{N}{r_2}\Phi(r_1-r_2)e^{-(2\bar{n}+1)\gamma^2(r_1-r_2)^2}=\sum_{r=-N}^N\binom{2N}{N+r}\Phi(r)e^{-(2\bar{n}+1)\gamma^2r^2}.$ The incoherent case is strikingly different, and we find $\mathcal{Q}=2^{2N}$.

\section{Wigner--Weyl quasiprobability distribution}
Describing our system in phase space makes subsequent calculations less cumbersome, so we choose to calculate the Wigner--Weyl quasiprobability distribution that corresponds to $\rho_\mathrm{mech}(0^+)$ derived above. To do so, we take the two-dimensional complex Fourier transform of the characteristic function of the state~\cite[Eqs.~(3.3.6) and (3.3.7)]{Scully1997}:
\begin{equation}
W(\beta_1,\beta_2)=\frac{1}{\pi^4}\iint\rmd^2b_1\,\rmd^2b_2\,e^{-2i\re{\beta_1b_1^\ast+\beta_2b_2^\ast}}\chi(b_1,b_2)\,,
\end{equation}
where $\chi(b_1,b_2)=\Tr\{\hat{D}_1(ib_1)\hat{D}_2(ib_2)\rho_\mathrm{mech}(0^+)\}$. Here, $\hat D_i(\alpha)$ ($i=1,2$) is the Weyl displacement operator introduced in the main text, operating on the mechanical state of mirror. A long but otherwise straightforward calculation yields
\begin{multline}
W(\beta_1,\beta_2)=\sum_{r_1,r_2=0}^N\binom{N}{r_1}\binom{N}{r_2}\frac{\Phi(r_1-r_2)e^{-\tfrac{2}{2\bar{n}+1}\bigl[\abs{\beta_1-\tfrac{i}{2}\gamma(2N-r_1-r_2)}^2+\abs{\beta_2-\tfrac{i}{2}\gamma(r_1+r_2)}^2\bigr]}}{\pi^2\bigl(\bar{n}+\tfrac{1}{2}\bigr)^2\mathcal{Q}}\\
\times\cos\bigl[2\gamma(r_1-r_2)\re{\beta_1-\beta_2}\bigr].
\end{multline}
This Wigner function is essentially a sum of Gaussian terms modulated by an interference factor (which is always equal to $1$ for the `classical' case). It can be written in a more symmetrical fashion by shifting it as $W_\mathrm{S}(\beta_1,\beta_2)\equiv W(\beta_1+\tfrac{i}{2}\gamma N,\beta_2+\tfrac{i}{2}\gamma N)$, whereby
\begin{multline}
W_\mathrm{S}(\beta_1,\beta_2)=\sum_{r_1,r_2=0}^N\binom{N}{r_1}\binom{N}{r_2}\frac{\Phi(r_1-r_2)e^{-\tfrac{2}{2\bar{n}+1}\bigl[\abs{\beta_1-\tfrac{i}{2}\gamma(N-r_1-r_2)}^2+\abs{\beta_2+\tfrac{i}{2}\gamma(N-r_1-r_2)}^2\bigr]}}{\pi^2\bigl(\bar{n}+\tfrac{1}{2}\bigr)^2\mathcal{Q}}\\
\times\cos\bigl[2\gamma(r_1-r_2)\re{\beta_1-\beta_2}\bigr]\,.
\end{multline}
Let us recall that a Wigner function with at least one pair of coordinates $(\beta_1,\beta_2)$ such that $W(\beta_1,\beta_2)<0$ does not have a classical explanation and necessarily describes a non-classical state. The measure of non-classicality used throughout the main text~\cite{Kenfack2004} is based on this observation, as discussed in Ref.~\cite{Kenfack2004}

\section{Joint homodyne detection}
Let us suppose we can read out any quadrature of the mechanical oscillator state through the cavity field. We would like to calculate the possibility of obtaining a joint measurement with amplitudes $x$ and $y$ and phases $\theta$ and $\phi$. Instead of reading out the individual cavity fields, however, we interfere them on a beamsplitter prior to homodyning. The beamsplitter operator, which we represent by $\hat{B}$ and define in the main text, effectively acts on the mechanical fields in the coherent representation as
\begin{equation}
\hat{B}\ket{\beta_1,\beta_2}=\ket{(\beta_1+i\beta_2)/\sqrt{2},(i\beta_1+\beta_2)/\sqrt{2}}\,.
\end{equation}
After operation by the beamsplitter, the state is projected onto the quadratures $x_\theta$ and $y_\phi$, and the absolute-squared value of the projection measured
\begin{equation}
\label{prob}
P(x_\theta,y_\phi)=\abs{\bra{x_\theta,y_\phi}\hat{B}\,\rho_\mathrm{mech}(0^+)\,\hat{B}^\dagger\ket{x_\theta,y_\phi}}^2\,.
\end{equation}
It is convenient to define the quantities
\begin{align}
\bar{x}_{\theta,r_1,r_2}&=\tfrac{1}{2}\gamma\bigl[(2N-r_1-r_2)\sin\theta-(r_1+r_2)\cos\theta-i(2\bar{n}+1)(r_1-r_2)(\cos\theta-\sin\theta)\bigr],\\
\bar{y}_{\phi,r_1,r_2}&=\tfrac{1}{2}\gamma\bigl[(r_1+r_2)\sin\phi-(2N-r_1-r_2)\cos\phi+i(2\bar{n}+1)(r_1-r_2)(\cos\phi-\sin\phi)\bigr],
\end{align}
so that the probability distribution in Eq.~\eqref{prob} becomes
\begin{multline}
P(x_\theta,y_\phi)=\frac{1}{\pi^2\bigl(\bar{n}+\tfrac{1}{2}\bigr)^2\mathcal{Q}^2}\Biggl|\sum_{r_1,r_2=0}^N\binom{N}{r_1}\binom{N}{r_2}\Phi(r_1-r_2)
e^{-\tfrac{1}{2\bar{n}+1}[(x_\theta-\bar{x}_{\theta,r_1,r_2})^2+(y_\phi-\bar{y}_{\phi,r_1,r_2})^2]}e^{-(2\bar{n}+1)\gamma^2(r_1-r_2)^2}\Biggr|^2.
\end{multline}
This expression is not quite the square of a sum of Gaussians because of the imaginary parts of $\bar{x}_{\theta,r_1,r_2}$ and $\bar{y}_{\phi,r_1,r_2}$ that, for the coherent case, lead to interference effects that are completely absent from the `classical' one. Indeed, if we decompose $\bar{x}_{\theta,r_1,r_2}=\bar{x}_{\theta,r_1,r_2}^\mathrm{r}+i\bar{x}_{\theta,r_1,r_2}^\mathrm{i}$, and $\bar{y}_{\phi,r_1,r_2}=\bar{y}_{\phi,r_1,r_2}^\mathrm{r}+i\bar{y}_{\phi,r_1,r_2}^\mathrm{i}$ we can rewrite this expression as
\begin{multline}
P(x_\theta,y_\phi)=\frac{1}{\pi^2\bigl(\bar{n}+\tfrac{1}{2}\bigr)^2\mathcal{Q}^2}\Biggl|\sum_{r_1,r_2=0}^N\binom{N}{r_1}\binom{N}{r_2}\Phi(r_1-r_2)
e^{-\tfrac{1}{2\bar{n}+1}[(x_\theta-\bar{x}_{\theta,r_1,r_2}^\mathrm{r})^2+(y_\phi-\bar{y}_{\phi,r_1,r_2}^\mathrm{r})^2]}\\
\times e^{i\sqrt{2}\gamma(r_1-r_2)\bigl[\cos\bigl(\theta+\tfrac{\pi}{4}\bigr)(x_\theta-\bar{x}_{\theta,r_1,r_2}^\mathrm{r})-\cos\bigl(\phi+\tfrac{\pi}{4}\bigr)(y_\phi-\bar{y}_{\phi,r_1,r_2}^\mathrm{r})\bigr]}\\
\times e^{-\bigl(\bar{n}+\tfrac{1}{2}\bigr)\gamma^2(r_1-r_2)^2\bigl[\sin^2\bigl(\theta+\tfrac{\pi}{4}\bigr)+\sin^2\bigl(\phi+\tfrac{\pi}{4}\bigr)\bigr]}\Biggr|^2\,.
\end{multline}

\subsection{Behaviour of resonance with number of particles}
Consider now $\re{\bar{x}_{\theta,r_1,r_2}}=\tfrac{1}{2}\gamma\bigl[(r_1+r_2)\sin\phi-(2N-r_1-r_2)\cos\phi\bigr]\,$, which is independent of $r_1$ and $r_2$ for $\sin\theta=-\cos\theta=\pm\tfrac{1}{\sqrt{2}}$. Let us further choose to monitor the coincidences for which $x=y$. In this case, when $\phi=-\theta-\tfrac{\pi}{2}$, we have $\re{\bar{x}_{\theta,r_1,r_2}}=\re{\bar{y}_{\phi,r_1,r_2}}$, and at $\theta=\theta_0=-\tfrac{\pi}{4}+2n\pi~(n\in\mathbb{N})$, all these curves (seen as functions of $\theta$ or $\phi$) coincide. At this point one observes interference effects in the quantum-mechanical case. After some algebra, we get to 
\begin{equation}
P(x_\theta,y_\phi)=\begin{cases}
\frac{1}{\pi^2(2\bar{n}+1)^2}&\quad\text{(incoherent~case)},\\
\frac{2^{4N}}{\pi^2(2\bar{n}+1)^2}\Bigl[\sum_{r=-N}^N\binom{2N}{N+r}e^{-(2\bar{n}+1)\gamma^2r^2}\Bigr]^{-2}&\quad\text{(coherent~case).}
\end{cases}
\end{equation}
The former of these equations is \emph{independent of the number of particles or the strength of the interaction}, whereas the latter is approximately linear in $N$. Indeed, for $N$ and $\gamma$ or $\bar{n}$ large enough we can write
\begin{equation}
P(x_\theta,y_\phi)=\begin{cases}
\frac{1}{\pi^2(2\bar{n}+1)^2}&\quad\text{(incoherent~case),}\\
\frac{N}{\pi(2\bar{n}+1)^2}&\quad\text{(coherent~case).}
\end{cases}
\end{equation}

\subsection{Effect of homodyning angle on the resonance peak}
With the choice of parameters analysed above, $P(x_\theta,y_\phi)$ has Gaussian peaks for both coherent and the incoherent cases. Let us now set $\theta=\theta_0+\nu$ and $\phi=-\theta_0-\tfrac{\pi}{2}+\nu$. It can be shown that, to lowest order in $\nu$, we have
\begin{equation}
P(x_\theta,y_\phi)=\begin{cases}
\frac{1}{\pi^2(2\bar{n}+1)^2}&\quad\text{(incoherent~case),}\\
\frac{1}{\pi^2(2\bar{n}+1)^2}\frac{\bigl|\sum_{r_1,r_2=0}^N\binom{N}{r_1}\binom{N}{r_2}e^{-2i\gamma^2(N-r_1-r_2)(r_1-r_2)\nu}\bigr|^2}{\bigl[\sum_{r=-N}^N\binom{2N}{N+r}e^{-(2\bar{n}+1)\gamma^2r^2}\bigr]^{-2}}&\quad\text{(coherent~case).}
\end{cases}
\end{equation}
There is again a qualitative difference between the two cases: The peak in the incoherent case is constant with respect to small variations in $\nu$, whereas the coherent case exhibits characteristic variations with such parameter.

\section{Momentum imparted by a colliding particle onto a mirror}
Let us work in a classical picture to estimate the effect of the adsorption of a single particle on a mirror. The change in (dimensionless) momentum of the mirror, $\Delta p$, due to the instantaneous adsorption of a particle travelling with momentum $p_\mathrm{part}$, is given by $M\omega_\mathrm{m}x_\mathrm{zpt}\Delta p=-p_\mathrm{part}$. Here $M$ is the mass of the mirror and $\omega_\mathrm{m}$ its oscillation frequency. We can write this as $\Delta p=-G$ with 
\begin{equation}
G=2\pi\times\frac{x_\mathrm{zpt}}{\lambda_\mathrm{dB}}=\frac{p_\mathrm{part}}{\sqrt{\hbar M\omega_\mathrm{m}}},
\end{equation}
where $\lambda_\mathrm{dB}=h/p_\mathrm{part}$ is the de Broglie wavelength of the matter wave. The extent of the zero-point fluctuations of the mirror motion is $x_\mathrm{zpt}=\sqrt{\hbar/(M\omega_\mathrm{m})}$. We divide $G$ by a factor of $\sqrt{2}$, in accordance with the definition of $\hat{D}_{1,2}$, to obtain the parameter $\gamma$ used in the paper.

\section{Frequency shift due to a single collision}
We work in a classical picture to approximate the frequency shift obtained in a resonant cavity as the result of the adsorption of a single particle. The momentum imparted onto the mirror results in a maximal displacement $\Delta L$ such that $M\omega_\mathrm{m}\Delta L=p_\mathrm{part}\Rightarrow\Delta L={p_\mathrm{part}}/(M\omega_\mathrm{m})$.
For a resonant cavity, we have $L_\mathrm{c}=\tfrac{n}{2}\lambda_\mathrm{c}$, where $n\in\mathbb{N}$. Thus,
\begin{equation}
\Delta L=\tfrac{n}{2}\Delta\lambda\Rightarrow\frac{\Delta L}{L_\mathrm{c}}=\frac{\Delta\lambda}{\lambda_\mathrm{c}}\,.
\end{equation}
We can therefore calculate the maximum frequency modulation
\begin{equation}
\omega_\mathrm{c}=\frac{2\pi c}{\lambda_\mathrm{c}}\Rightarrow\Delta\omega=\frac{\omega_\mathrm{c}}{L_\mathrm{c}}\lvert\Delta L\rvert=\frac{\omega_\mathrm{c}p_\mathrm{part}}{L_\mathrm{c}\omega_\mathrm{m}M}=2\pi\times\frac{\omega_\mathrm{c}x_\mathrm{zpt}^2}{L_\mathrm{c}\lambda_\mathrm{dB}}\,.
\end{equation}
\end{widetext}

\end{document}